\documentclass[twocolumn,prd,floatfix]{revtex4}
\usepackage{color}
\usepackage{tabularx}
\usepackage{epsfig}
\usepackage{amsmath}
\usepackage{amssymb}
\usepackage{bm}
\usepackage{graphicx}
\usepackage{multirow}

\usepackage{epsfig}

\begin{document}

\title{Investigation of two colliding solitonic cores in Fuzzy Dark Matter models}

\author{Alireza Maleki, Shant Baghram and Sohrab Rahvar}
\affiliation {Department of Physics, Sharif University of Technology, P. O. Box 11155-9161, Tehran, Iran.}

\date{\today}

\begin{abstract}
One of the challenging questions in cosmology is the nature of  dark matter particles.
Fuzzy Dark Matter (FDM) is one of the candidates which is made of very light  ($m_{FDM}\simeq 10^{-22}-10^{-21}$ eV) bosonic particles  with no self-interaction. It is introduced by the motivation to solve the core-cusp problem in the galactic halos.
In this work, we investigate the observational features from FDM  halo collisions. Taking into account the quantum wave-length of the condensed bosonic structure, we determine the interference of the wave function of cores  after collision. The fringe formation in the wave function is associated to the density contrast of the dark matter inside the colliding galaxies.  The observational signatures of the fringes of the distribution of the dark matter are  (i) on the  lensing of the background sources, (ii) accumulation of the baryonic plasma tracking the interference of the FDM potential and (iii) excess in the X-ray emission from dense regions. Finally, we provide prospects for the observations of quantum wave features of FDM in the colliding galaxies. The NGC6240 colliding galaxy at the redshift of $z=0.024$  is a suitable candidate for this study.
No signal is detected from the fringes in the Chandra data and taking into account the angular resolution of 
the  telescope, we  put constrain of $m> 7 \times10^{-23}$ eV on the mass of FDM particles.
\end{abstract}

\maketitle

\section{introduction}

According to the standard cosmological $\Lambda$-Cold Dark  Matter ($\Lambda$CDM) model, the Universe is made of about $68\%$ dark energy and $28\%$ dark matter with $4\%$  baryonic matter \cite{aghanim2018planck}. The largest contribution of the energy-momentum in the Universe is believed to be dark with an unknown nature. In the $\Lambda$CDM model, dark matter particles are modeled as the cold collisionless particles \cite{bernabei2003dark,markevitch2004direct}. That is remarkably successful in describing the Universe in the large scales from cosmic microwave background radiation to  the formation of cosmic structures \cite{springel2005simulations}.
However, it seems that $\Lambda$CDM faces with many serious problems and tensions in comparison with the observational data  in the small scales. One of the problems is  that  the CDM simulations lead to the singular density at the center, varying as $r^{-1}$, however the observational data indicate the smoothly varying core density. This  inconsistency is called the core-cusp problem  \cite{alvarado1993observational,moore1999cold,weinberg2015cold}. The other tensions of the standard model in the small scales are the missing satellites problem \cite{klypin1999missing}, too big to fail problem \cite{boylan2011too}, alignment of the substructures in a galactic halo \cite{schneider2012shapes}.  Beside that, all the attempts  for the detection of the dark matter particles till now are not successful \cite{aprile2017first}. Accordingly, there are so many proposed models and suggestions developed in recent years in order to solve the CDM non-detection problem in one hand and to conceal the small scale crisis on the other hand \cite{avila2001formation,rocha2013cosmological,governato2010bulgeless,garrison2013can}.

One of the hypotheses that have taken attention recently as a new model to solve the CDM problems, assumes that dark matter is composed of very light Bosonic particles, forming Bose-Einstein Condensate (BEC) core \cite{hu2000fuzzy}.
Apart from its vast role in the condense matter physics, the BEC is thought to be an important concept in cosmology and astrophysics
\footnote{For example Boson stars  might be formed from the condensation of scalar fields\cite{kaup1968klein,ruffini1969systems,schunck2003general}.
 It has been showed that neutron stars also could form Cooper pairs and behave as boson stars \cite{abuki2007bcs,chavanis2012bose}}. \\  In this paper, invoking the  physics of BEC, it  provides a model for solving some of the problems of the dark matter physics  in the small scales, where dark matter (DM) behaves as a coherent wave function \cite{hu2000fuzzy,bohmer2007can,woo2009high}.
One class of these models, assumes that the dark matter consists of very light particles with mass range of ($m\simeq 10^{-22}-10^{-21}$ eV) \cite{hui2017ultralight}, having the de Brogli wavelength of the order of {\it kpc} where the quantum mechanical behavior are important in these scales. This model often called Fuzzy Dark Matter (FDM) \footnote{There are alternative names such as 'Wave Dark Matter', 'Ultra-light Axionic Particles (ULP)', 'BEC DM', and so on \cite{lee2018brief}}.

  One of the candidates for this type of dark matter particles is the Axion-like particles. 
 This light bosonic particles seems to be well motivated in high energy particle physics  \cite{marsh2016axion}. They are also called  Axion-like particles because one of them could be a candidate for the QCD Axion. Also, all the models of the string theory have at least several bosonic fields of such particles \cite{hui2017ultralight}.

In this scenario, taking light bosons as the dark matter particles, at the large scales, the predictions are the same as the CDM. However because of  the quantum mechanical effects, it has different predictions in the small scales. This model can be taken potentially a good candidate for the dark matter, consistent with the observational data both in the large and in the small scales \cite{schive2014understanding}. One of the features  is the quantum pressure where in the gravitational collapsing systems such as galaxies stabilize the system and prevent the formation of cusp at the center of the halo.
In the FDM hypothesis, the dark matter halo consists of a solitonic core that formed BEC at the inner region of the halo and a non-condensed dark matter particles at the outer region of the halo with a NFW profile \cite{schive2014understanding,   navarro1995simulations,du2016substructure}. Although it seems that FDM candidate solves many issues of DM, however some specific observations are necessary to verify this type of dark matter.

 In this work, we investigate the observational features of FDM in the interference formation from the
colliding galaxies. We take into account the collision of the two solitonic waves as the core of two colliding galaxies.  The wave-like nature of the core can affect the dynamics of the halos in collision and it can be observed in the various features of a gravitating soliton-soliton collision. In this direction, there are some simulations in one-dimensional FDM collisions \cite{bernal2006scalar,gonzalez2011interference,veltmaat2016cosmological,schwabe2016simulations} .
Due to the interference, a pattern in FDM density field with the corresponding gravitational potential is generated. While this structure is dark and difficult to detect, the gravitational potential-fringes can be a potentially prominent observable, playing the role of the gravitational lensing to deflect the light of the distant cosmological sources. The effect would be both traced by shear and the convergence. In this work, we calculate the strength of the gravitational lensing by the colliding galaxies. Also, we investigate the gravitational interaction of baryonic matter with the  interference pattern gravitational potential of the FDM. For  two dissipative and non-dissipative regimes of the baryonic matter, depending on the cooling time-scale, we calculate the dynamics of the  baryonic matter profile within the gravitational potential of FDM. Due to heating mechanism, the baryonic matter form hot plasma during the collision of the two galaxies and we expect to observe the ripples in
the plasma profile from the X-ray measurements.

The structure of this work is as follows: In Section (\ref{S2}) we calculate the  mass density profile and column density of DM after collision of the two
halo cores. In Section (\ref{S3}) we calculate the lensing effect due to the fringes from the collision of the galactic halos. Also, we derive the shear and the convergence signals from the lensing. In Section (\ref{density}), we calculate the ripples in the profile of  baryonic matter, resulting from the gravitational interaction of baryonic matter with the interference pattern of the dark matter. Also, we investigate the intensity of the X-ray radiation from baryonic matter. Finally in Section (\ref{ObserConsrains}), we address the  comparing of  the theoretical predictions to  the observational data. The conclusion is given in Section (\ref{conc}).

\section{collision of the two FDM solitonic halo cores }
\label{S2}
 In this section we study the theoretical background of the FDM and two halo core collision in this theory. In the non-relativistic regime, we study the dynamics of a coherent gravitational self-interacting massive scalar field.
 The dynamics of the wave function $\psi$, representing this field follows the Schrodinger-Poisson (SP) equations as \cite{schwabe2016simulations}
 \begin{align}
 i\hbar \frac{\partial \psi}{\partial t} =&-\frac{\hbar^{2}}{2m} \nabla^2\psi+mU\psi, \\
  \nabla^2 U=&4\pi G\rho,
 \label{psequation}
 \end{align}
where $\psi$ is the wave function, $U$ is the gravitation potential of the system and $m$ is the mass of the FDM particles. In this equation the density is defined by $\rho={\mid \psi ^{2}\mid}$. One can use the Madelung transformation \cite{madelung1927quantentheorie}
with writing the wavefuction as $\psi(\vec{r},t)=\sqrt{ \rho(\vec{r})}\exp(i{S(\vec{r},t)}/{\hbar})$, in which we can read the velocity from $\bm{v}=\bm{\nabla} S/m$. If we substitute this wave-function in the SP equations, separate the imaginary and real parts we obtain \cite{chavanis2011mass}

\begin{align}
\frac{\partial \rho}{\partial t}+\bm{\nabla}.(\rho \bm{v})=0 ,
\label{qeuler} \\
\frac{\partial \bm{v}}{\partial t}+(\bm{v}.\nabla).\bm{v}+\nabla{U}+\frac{\nabla{Q}}{m}=0.
\label{qeulerlike}
\end{align}
With the Poisson equation, this is an equivalent hydrodynamic description of the SP equations. Here, $Q$ is the quantum potential which defines as
\begin{align}
Q=-\frac{\hbar^2}{2m}\frac{\nabla^2{\sqrt{ \rho}}}{\sqrt{\rho}}.
\label{qunatumpressure}
\end{align}
This term causes a pressure known as a quantum mechanical pressure.
The gravitational interactions tend to contract the system and on the other hand, the quantum mechanical pressure has a repulsive effect on the wave function and causes the wavefunction broadening. In the stationary state, these two effects stabilize the system and make a soliton core. We define the $r_{c}$ as the core radius, where the density reduces to half of its maximum value. This radius can be obtained from numerical solution of  equation (\ref{qeulerlike}) and the mass enclosed in this radius can be considered as the core mass 
\begin{align}
M_{r_{c}}\simeq 1.54\frac{\hbar^2}{Gm^2 r_{c}},
\label{mrc}
\end{align}
where for the large mass of the solitonic core, the corresponding radius  is smaller.
Simulation based on these equations for FDM  in the comoving frame is applied to investigate the structure formations in universe filled with FDM, where the fitting function for density from simulations is as below \cite{schive2014understanding,schwabe2016simulations}
\begin{align}
 \rho_{c} (r)\simeq \rho_{0}[1+0.091(\frac{r}{r_{c}})^2]^{-8},
 \label{cdensity profile}
 \end{align}
where $\rho_{c}(r)$ is the density of condensate state and $\rho_{0}$ which known as the the  core's central mass density is given  by
 \begin{align}
 \rho_{0}  \simeq 1.9 \times(\frac{10^{-23}\text{eV}}{m})^{2}(\frac{\text{kpc}}{r_{c}})^{4} M_{\odot}pc^{-3}.
 \label{cdensity0}
 \end{align}
 Also, the core size and the halo mass ($M_{h}$) has the following relation
  \begin{align}
 r_{c}\simeq160 (\frac{M_{h}}{10^{12} M_\odot})^{-\frac{1}{3}}(\frac{m}{10^{-22}ev})^{-1} pc,
 \label{crdensity0}
 \end{align}
 in which we normalized the halo mass to the mass of Milky Way and the FDM particle mass is normalized to $10^{-22}$~eV. 
 In this work for simplicity  we  take the approximation of constant density for the solitonic core. Unlike to the center, the outer region of FDM halo  behaves like the NFW profile \cite{schive2014understanding}. In other word, the halo profile would consist of two regions of the core at the center and NFW at the outer region. The transition between these  two phases happens in  some distance larger than  the core size \citep{robles2018scalar}, we  call this transition distance  $r_t$. Then, the  mass density of halo could be written  as follows \cite{du2016substructure}:
 \begin{align}
 \rho(r) =\rho_{c}(r)\Theta(r_t-r)+\rho_{NFW}(r)\Theta(r-r_t),
 \end{align}
where the $\Theta$ is the Heaviside function and $\rho_{NFW}(r)$ is the NFW profile for the outer parts of halo.

Now we take the collision of two solitonic halo cores in the FDM model.
The general analysis of this problem needs to take into account the mutual gravitational interaction of the two halos and the effective quantum mechanical pressure in the dynamics of the system according to the SP equations.  Simulations which investigate the collision of the two solitonic cores shows that depending on the velocity of the collision we can have multiple interference fringe in the mass density \cite{bernal2006scalar,gonzalez2011interference,veltmaat2016cosmological,schwabe2016simulations,navarrete2017spatial}. As a result they showed that
 if the system is unbounded, in other words, the total energy of the system is positive  the two solitons which  pass through each other  would not experience  significant change in their profiles. On the other hand, if the two solitons with the core of size $r_c$, merges and form a bound system,   after relaxation it form a stationary state and the stability length from equation (\ref{mrc}) decrease to ${r_c}/{2}$.

In this work we investigate the unbounded collisions in which during the collision of the two solitons, the superposition of the two waves produces the interference pattern which results in interference pattern in the mass density profile of the two systems.
 For simplicity we take a flat
wave for each soliton with a cut-off to confine these waves.
Then from the superposition of the two waves, the overall wave function is:
\begin{multline}
 \label{psequation}
 \psi(\vec{r},t;\vec{r_1},\vec{r_2}) =\sqrt{ \rho_{1}(\vec{r}-\vec{ r_{1}})}e^{i(\vec{k_{1}}.\vec{r}+\phi +\omega t)}\\
+\sqrt{ \rho_{2}(\vec{r}-\vec{ r_{2}})}e^{i(\vec{k_{2}}.\vec{r}+\omega t)},
 \end{multline}
where $\psi$ is a function of coordinate $\vec{r} = (x,y,z)$ and time and also  $\vec{r}_{1}$ and  $\vec{r}_{2}$ which are the positions of the centers of the two solitons with respect to their  center of mass,
$\rho_{1}(\vec{r}-\vec{ r_{1}})$ and  $\rho_{2}(\vec{r}-\vec{ r_{2}})$ are the density profile of the solitons,  $\vec{k}_{1}$ and  $\vec{k}_{2}$  are the wave-numbers corresponding to each soliton and $\phi$ is their relative phase difference.
We take a top-hat profile for the density of solitons. The overall density would be $\rho(\vec{r},t) =  \psi(\vec{r},t)  \psi(\vec{r},t)^\star$ which results is

\begin{figure}
\centering
\includegraphics[scale=0.5]{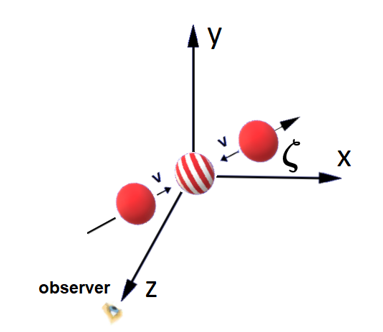}
\caption{Schematic figure from the interference of two halo solitonic cores. The z-direction is set to be along the line of sight. We plotted a head on collision in which the angel between the x-axis and the line of collision is $\zeta$.}
\label{fig:Entropyplot}
\end{figure}


\begin{eqnarray}
&&\rho_{t}(\vec{r}, t;\vec{r_1},\vec{r_2} )=\bar{\rho}_{1}\Theta(|\vec{r}-\vec{r_1}|-R_1)+\bar{\rho}_{2}\Theta(|\vec{r}-\vec{r_2}|-R_2)\nonumber \\
&+&2\sqrt{\bar{\rho}_{1}\bar{\rho}_{2}\Theta(|\vec{r}-\vec{r_1}|-R_1)\Theta(|\vec{r}-\vec{r_2}|-R_2) } \nonumber \\
&\times&\cos((\vec{k_{1}}-\vec{k_{2}}) . \vec{r}+ \phi),
\label{rointerf1}
\end{eqnarray}
where $\bar{\rho}_{1}$  and $\bar{\rho}_{2}$ are the constant density of solitons within the radius of $R_1$, $R_2$. 
As we discussed before, we assume that the colliding solitons are in an unbound state and they are in their primier encounter. Accordingly the gravitational interaction of them do not change the size of cores, as the relaxation time is much larger than the encounter time. This assumption is supported by the solitonic collision simulations  \cite{bernal2006scalar,gonzalez2011interference,veltmaat2016cosmological,schwabe2016simulations,navarrete2017spatial}. 

Since the unperturbed solitons follows the step function profile, the first two terms in Eq. (\ref{rointerf1}) provide constant terms while the third term results from the the interference of the two waves. Analogue to optics, we can define a parameter which shows the visibility of the interference as $\nu=({\bar{\rho}_{tmax}-\bar{\rho}_{tmin}})/({\bar{\rho}_{tmax}+\bar{\rho}_{tmin}})$, in which the index $max$ and $min$ indicates the maximum and minimum value of the total mass density.  In our case from equation \eqref{rointerf1} we get $\nu={2\sqrt{\bar{\rho}_{1}\bar{\rho}_{2}}}/({\bar{\rho}_{1}+\bar{\rho}_{2}})$, this parameter changes from zero to one. For the case when two solitons mass density are equal, the interference visibility would be maximum.

In what follows we calculate the column density of the two colliding solitons. We will use
this parameter for studying the gravitational lensing from this system.
\begin{figure}
\centering
\includegraphics[scale=0.35]{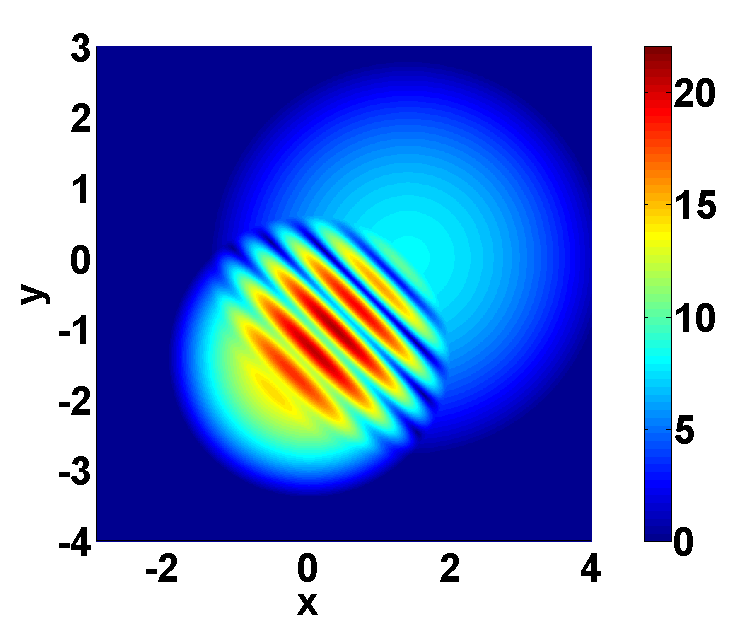}
\caption{A snapshot of the  interference of two halo solitonic cores. The larger halo has the radius of $\sqrt{2}$ times larger than the smaller halo(which has a radius of $ 2$) . The wavenumber for the small halo is $k\hat{y}$, and the larger soliton, that is $-k\hat{x}$ with equal size of $k=10$. We normalized the mass density to the mass density of the smaller soliton. The $x$ and $y$ axis are in arbitrary units.}
\label{solitonoverlap}
\end{figure}
For simplicity in our calculations, we use the polar cordinate $\vec{\xi'}= \xi'(\cos(a),\sin(a))$,  where $\xi'^2 = x^2+y^2 $ and  $a$ is the angle of $\xi'$ with respect to $x$ axis 
 .
 For the colliding solitons, the vector of wave number can be decomposed in the $(x,y)$ plane and $z$ axis as
 $k_{1}=k_{1\xi'}\hat{\xi'_{1}}+k_{1z}\hat{z}$ and $k_{2}=k_{2\xi'}\hat{\xi'_{2}}+k_{2z}\hat{z}$ where $\hat{\xi'}$ represents the unit vector along $\xi'$. 

 The column density of the merging system in this new coordinate system is given as
\begin{eqnarray}
 \label{sigintcompelet}
\Sigma (\vec{\xi'})&=& \int_{-z}^z\rho (x,y,z) dz   \\ \nonumber
&=&  \Sigma_0(\xi')  + 4\frac{\sqrt{\bar{\rho}_{1}\bar{\rho}_{2}} \cos(k_{1\xi'}\xi'_{1}-k_{2\xi'}\xi'_{2}+\phi)}{\Delta k_{z}}\times \\ \nonumber
 &&\sin( \Delta k_{z} (\sqrt{R^2-\xi'^{2}}))\Theta(r_{1\xi'}-R_1)\Theta(r_{2\xi'}-R_2),
\end{eqnarray}
in which $\Delta k_{z}=k_{1z}-k_{2z}$, and $r_{1\xi'}$ and  $r_{2\xi'}$ are the positions of two soliton in the plane of projection to our line of sight.
For simplicity in the form of equations we define $\Sigma_0 = 2(\bar{\rho}_{1}\Theta(r_{1\xi'}-R_1)+\bar{\rho}_{2}\Theta(r_{2\xi'}-R_2))\sqrt{R^2-\xi'^{2}}$ as the background column density in Eq. (\ref{sigintcompelet}). The second term in this equation results from the interference between the two solitons.
The non-perpendicular direction for the relative velocity of the two solitons with respect to the line of sight results in a combination of the
linear and the circular pattern of the interference, that results from the equation \eqref{sigintcompelet}. The linear part results from the component of wavenumber perpendicular to  the observer. This pattern in $x-y$ plane has slope of $-({k_{x2}-k_{x1}})/({k_{y2}-k_{y1}}), $(see Figure \eqref{solitonoverlap} ).
It is worth to mention that the orientation of the fringes depends on the direction of the initial $k$-vectors and the width of the fringes depend on their relative velocities. 

For the head-on collisions, we  set the line of sight in the z-direction and the line of collision (i.e. $\vec{n} = \hat{k}_1 - \hat{k}_2$) to be in x-z plane.  This is equivalent to choosing a coordinate where $y$ axis components for wavenumber vectors (see Fig. \ref{fig:Entropyplot}) set to zero in Eq.\eqref{sigintcompelet}, accordingly we will have the column density as follows
\begin{eqnarray}
 \label{sigcom1}
&& \Sigma (\vec{\xi'})=\Sigma_0
 + 4\frac{\sqrt{\bar{\rho}_{1}\bar{\rho}_{2}}}{k\sin(\zeta)}\sin[k\sin(\zeta)\sqrt{R^2-\xi'^{2}}]  \\
&\times&\cos[k\xi'\cos(\zeta)\cos(a)+\phi] \Theta(r_{1\xi'}-R_1)\Theta(r_{2\xi'}-R_2), \nonumber
\end{eqnarray}
in which the $\zeta$ is the angle between the line of collision and the $x$ axis.
In the next section, we will study the gravitational lensing of two fuzzy dark matter core collision.

\section{gravitational lensing of colliding two FDM cores}
\label{S3}
The gravitational lensing of the two FDM core collision is important from the observational point of view. Since the FDM has no electromagnetic signal, we may prove its existence from the gravitational lensing effect. The most important signal of FDM in the colliding galaxies would be the detection of the interference pattern which does not happen to the ordinary dark matter.

 In order to calculate the lensing effect for the interference mass density profile, we assume the thin lens approximation
 which is applicable for this system. The deflection of a light ray crossing projected  mass distributed  $\Sigma (\vec{\xi})$  at $\xi$  is given by \cite{schneider2006gravitational}.
 \begin{equation}
\hat{\alpha}_d(\vec{\xi})=\frac{4G}{c^2}\int\frac{(\vec{\xi}-\vec{\xi'})\Sigma(\vec{\xi'})}{\mid \vec{\xi}-\vec{\xi'} \mid ^{2}} d^{2}\xi'.
\label{denangge}
 \end{equation}
\begin{figure}
\centering
\includegraphics[scale=0.3]{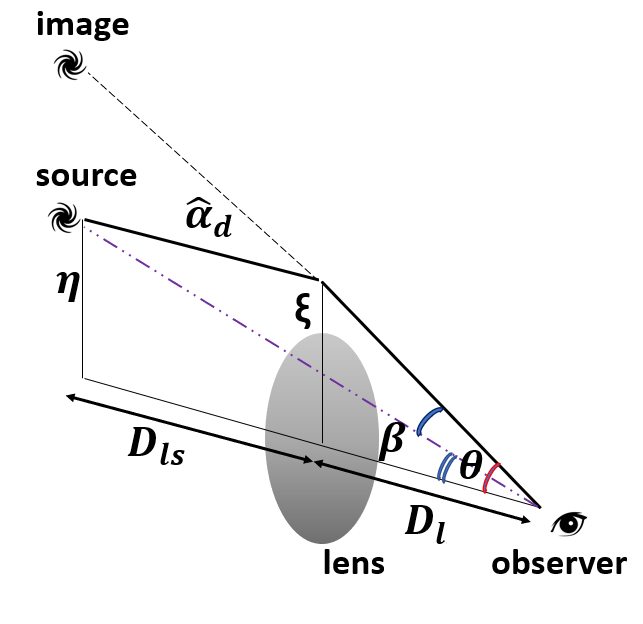}
\caption{A schematic view of a thin gravitational lens system. $D_l$, $D_s$ and $D_{ls}$ are the distances to  lens, source, lens-source, respectively. Also, $\xi$  and $\eta$ are the physical distance of images and  source in the order.  $\alpha_d$ is the deflection angel, $\beta$ is the angular position of the source in the absence of the lens and $\theta$ is the angular position of image and $\alpha= \theta - \beta$.}
\label{gravlesngal}
\end{figure}
In Figure (\ref{gravlesngal}) a typical lensing situation is depicted. A lens located at the distance of $D_{l}$ and the source at distance $D_{s}$ from the observer and the distance between the lens and the sources is $D_{ls}$. The lensing equation is given by
    \begin{equation}
    \eta=\xi \frac{D_{s}}{D_{l}}- \hat{\alpha}_d(\vec{\xi}) D_{ls},
    \label{lensequ1}
 \end{equation}
 where $\xi$ is the physical distance of images and $\eta$ is the distance of the source from our line of sight to the position of lens. For simplicity, we use the following coordinates of  $\eta=D_{s}\beta$ , $\xi=D_{l}\theta$. The lens equation is rewritten as
  \begin{equation}
   \vec{\beta}=\vec{\theta}- \frac{D_{ls}}{D_{s}} \vec{\hat{\alpha}}_d .
 \end{equation}
 It is convenient to write this equation in dimensionless form. So, by defining $\vec{\chi}={\vec{\xi}}/{R}$, in which $R$ is the actual size of the lens 
 and $\vec{\Upsilon}={\vec{\eta}}/{\eta_{0}}$
where in this definition  $\eta_{0}=R{D_{s}}/{D_{l}}$, accordingly the  equation \eqref{lensequ1} simplifies to
   \begin{equation}
    \vec{\Upsilon}=\vec{\chi} -\vec{\alpha}_d
    \label{lensequ2xy}
 \end{equation}
where we defined $\vec{\alpha}_d=\frac{D_{ls}D_{l}}{D_{s}R}\vec{\hat{\alpha}}_d$. We can also define the convergence as below
  \begin{equation}
  \kappa(\vec{\theta})=\frac{\Sigma(D_{l}\vec{\theta}) }{\Sigma_{cr}},
 \end{equation}
 where $\Sigma_{cr}$ is the critical mass density and defined as:
 \begin{equation}
\Sigma_{cr}=\frac{c^{2}}{4\pi G}\frac{D_{s}}{D_{ls}D_{l}}.
 \end{equation}
The deflection potential related to the $\kappa$ throughout
 \begin{equation}
 \nabla^2 \Psi(\vec{\chi})=2\kappa(\vec{\chi}),
 \label{defpottoalpha}
 \end{equation}
 where $\nabla $ is derivative with respect to cosmological comoving distance.
 It also related to the deflection of light $ \nabla \Psi(\vec{\chi})=\vec{\alpha}_d(\vec{\chi})$.
Now by using the Green function in two dimension, the deflection potential is
 \begin{equation}
  \label{psi}
   \Psi(\vec{\chi})=\frac{1}{\pi} \int \kappa(\vec{\chi'}) \ln{{\mid \vec{\chi}-\vec{\chi'} \mid}}d^{2}\chi'.
 \end{equation}
 In order to study the effect of FDM on the lensing convergence,  we use the definition of the column density from the collision of the two cores in Eq. \eqref{sigcom1}, then the $\kappa$  is given by
\begin{multline}
\kappa(\vec{\chi'})=  \frac{4\pi G}{c^{2}}\frac{D_{ls}D_{l}}{D_{s}}(2(\bar{\rho}_{1}+\bar{\rho}_{2})R\sqrt{1-\chi'^{2}} + \\ \frac{4\sqrt{\bar{\rho}_{1}\bar{\rho}_{2}}}{k\sin(\zeta)} \sin[k\sin(\zeta)R\sqrt{1-\chi'^{2}}] \\ \times \cos[kR\chi'\cos(a)\cos(\zeta)+\phi]),
\label{kappa}
\end{multline}
where integrating Eq. (\ref{psi}) for a simple case assuming the two colliding halos have the same mass, we get the result of
\begin{equation}
\label{factor}
 \Psi(\vec{\chi};R,k,\phi)=\frac{16 GR}{c^{2}}\frac{D_{ls}D_{l}\bar{\rho}}{D_{s}} F(\chi,b,\zeta, k, \phi)
 \end{equation}
 where the dimensionless function of $F$ is given by
\begin{equation}
F(\chi,b,\zeta, k, \phi)= RI_{1}+\frac{I_{2}}{k\sin(\zeta)},
\end{equation}
in which
\begin{eqnarray}
I_{1} &=&
\int_{0}^{1}\int_{0}^{2\pi}  \chi'd\chi' da  \sqrt{1-\chi'^{2}}  \nonumber \\
&&  \ln \left(\sqrt{\chi'^{2}+\chi^{2}-2\chi \chi' \cos(a-b)} \right), \nonumber \\
I_{2} &=&
  \int_{0}^{1}\int_{0}^{2\pi} \chi'd\chi' da  \ln \left(\sqrt{\chi'^{2}+\chi^{2}-2\chi \chi' \cos(a-b)} \right) \nonumber \\
  && \left( \sin \left[k\sin(\zeta)\sqrt{1-\chi'^{2}} \right]\cos \left[kR\chi'\cos(a)\cos(\zeta)+\phi \right] \right). \nonumber \\
\end{eqnarray}
In what follows, for simplicity we will calculate the dimensionless $F$ function. The lensing potential obtain by multiply it to the corresponding factor  $ {\mathcal{K}}=16 G R D_{ls}D_{l}\bar{\rho}/(c^{2}D_{s})$ in Eq. (\ref{factor}).

\begin{figure}
\centering
\includegraphics[scale=.35]{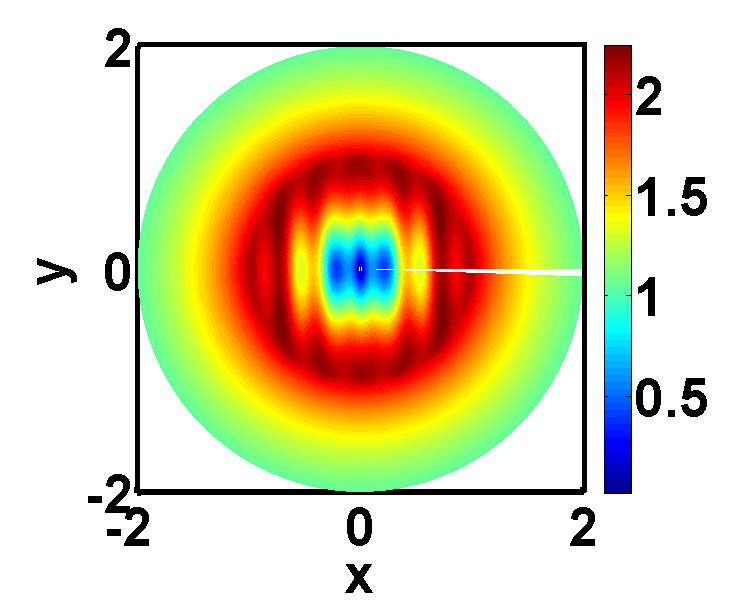}
\caption{The contour plot of the deflection angle due to the collision of two solitons is plotted. The deflection angle, approximately changes proportional to the inverse of the distance, outside the solitonic core. We normalized the deflection angle to $ {\mathcal{K}}=16 G R D_{ls}D_{l}\bar{\rho}/(c^{2}D_{s})$.The $x$ and $y$ axis are in arbitrary units. The wavenumber is normalized to the characteristic  wavenumber which is  defined as $k^{\star}=1/r_c$ and we set $k=20$,  assuming core size $r_c=1$.}
\label{deflectonang}
\end{figure}

 Figure \eqref{deflectonang} represents the  deflection angle ($\hat{\alpha}_d$) for a lens  results from the collision of the two solitons where at the inner region there is an interference pattern. As we expected  the deflection decrease with the inverse of the scale outside of the lens.  Deflection angle can be measured from \eqref{defpottoalpha}.
 Also the distortion of  the image is given by the Jacobean of the equation \eqref{lensequ2xy} as follows
  \begin{equation}
    A=\frac{\partial \vec{\Upsilon}}{\partial \vec{\chi}}.
 \end{equation}
Here is the distortion matrix as follows
 \begin{equation}
 A=\delta_{ij}-\frac{\partial^2\Psi(\vec{\chi})}{\partial \chi_{i}\partial \chi_{j}}=
  \begin{bmatrix}
    1-\kappa-\gamma_{1} & -\gamma_{2} \\
   -\gamma_{2} & 1-\kappa+\gamma_{1},
  \end{bmatrix}
  \end{equation}
 where we adapt the following conventional notation of  $\frac{\partial^2\Psi(\vec{\chi})}{\partial \chi_{i}\partial \chi_{j}}=\Psi(\vec{\chi})_{ij}$, accordingly  $\gamma_{1}(\vec{\chi})=\frac{1}{2}(\Psi_{11}-\Psi_{22})$ and $\gamma_{2}(\vec{\chi})=\Psi_{12}=\Psi_{21}$. Then martix $A$ can be written in a simpler form of
  \begin{equation}
 A=(1-\kappa) \begin{bmatrix}
    1 & 0 \\
   0& 1
  \end{bmatrix}
   +
   \begin{bmatrix}
    -\gamma_{1} & -\gamma_{2} \\
   -\gamma_{2} & \gamma_{1}
  \end{bmatrix}.
   \end{equation}
 The first term which induces an isotropic distortion results from the convergence alone, the second term so-called shear  stretches the image along two directions. We can write the shear matrix in a coordinate rotated by an angle $\delta$ such that 
  \begin{equation}
   \begin{bmatrix}
    -\gamma_{1} & -\gamma_{2} \\
   -\gamma_{2} & \gamma_{1}
  \end{bmatrix}
  =
 -\gamma_{t} \begin{bmatrix}
    \cos(2\delta) & \sin(2\delta) \\
  \sin(2\delta) &-\cos(2\delta)
  \end{bmatrix},
  \end{equation}
  Where $\gamma_{t}=\sqrt{\gamma_{1}^2+\gamma_{2}^2}$  is the magnitude of the shear, and $\delta$ describes it's rotation orientation. The magnification factor, $\mu$ of the lens can be compute by:
  \begin{equation}
 \mu=\frac{1}{det(A)}.
 \label{mumagnific}
 \end{equation}

For small value of the convergence we will have $\mu={1}/{det(A)} \simeq 1+2\kappa$. In what follows we will show how the collision of solitonic cores can introduce the lensing effect both in weak and strong regimes.

 \begin{figure}
\centering
\includegraphics[scale=.22]{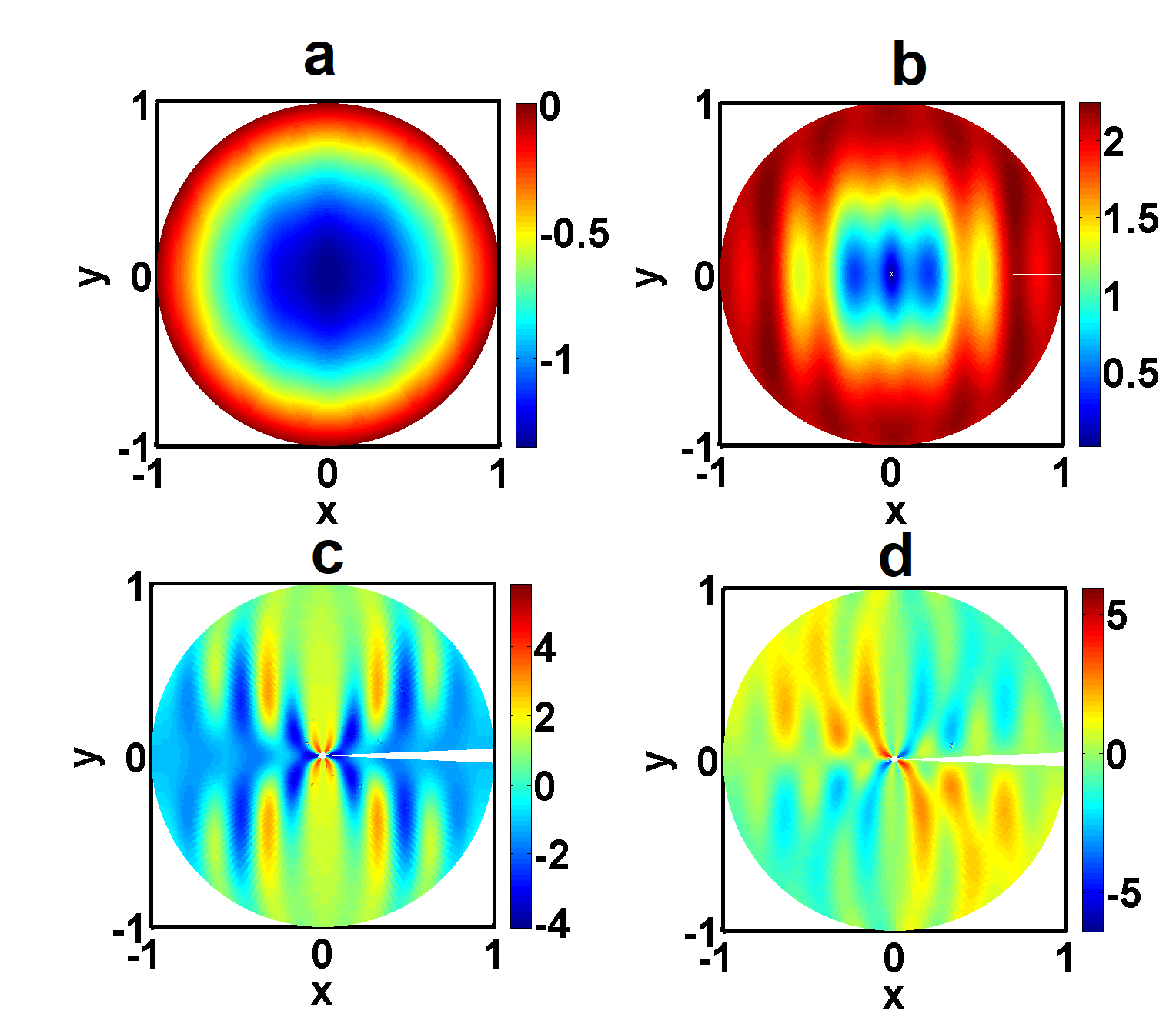}
\caption{The lensing parameters calculated for  two solitonic interference lens. The  phase of collision is adopted to ($\phi=0$). The observational angle  that is the complementary of angle between the observer and the line of collision  also is chosen to be  zero. In other words  perpendicular to the line of collision. The panels are as follows (a) deflection potential (b) deflection angle, (c) $\gamma_1$ the shear parameter (d)  $\gamma_{2}$, the second shear parameter. The $x$ and $y$ axis are in arbitrary units. The wavenumber is normalized to the characteristic  wavenumber which is  defined as $k^{\star}=1/r_c$ and we set $k=20$ with the assuming core size $r_c=1$.}
\label{pic1}
  \end{figure}
Figure (\ref{pic1}) shows the deflection potential and deflection angle with shear parameters for the  
case of collision phase of $\phi=0$ . In all the Figures, we have omitted the constant coefficients and worked with the normalized deflection potential. 
 Assuming the lens size is equal to core radius, $\mathcal{K}$ can be written as:
\begin{figure}
	\centering
	\includegraphics[scale=.2]{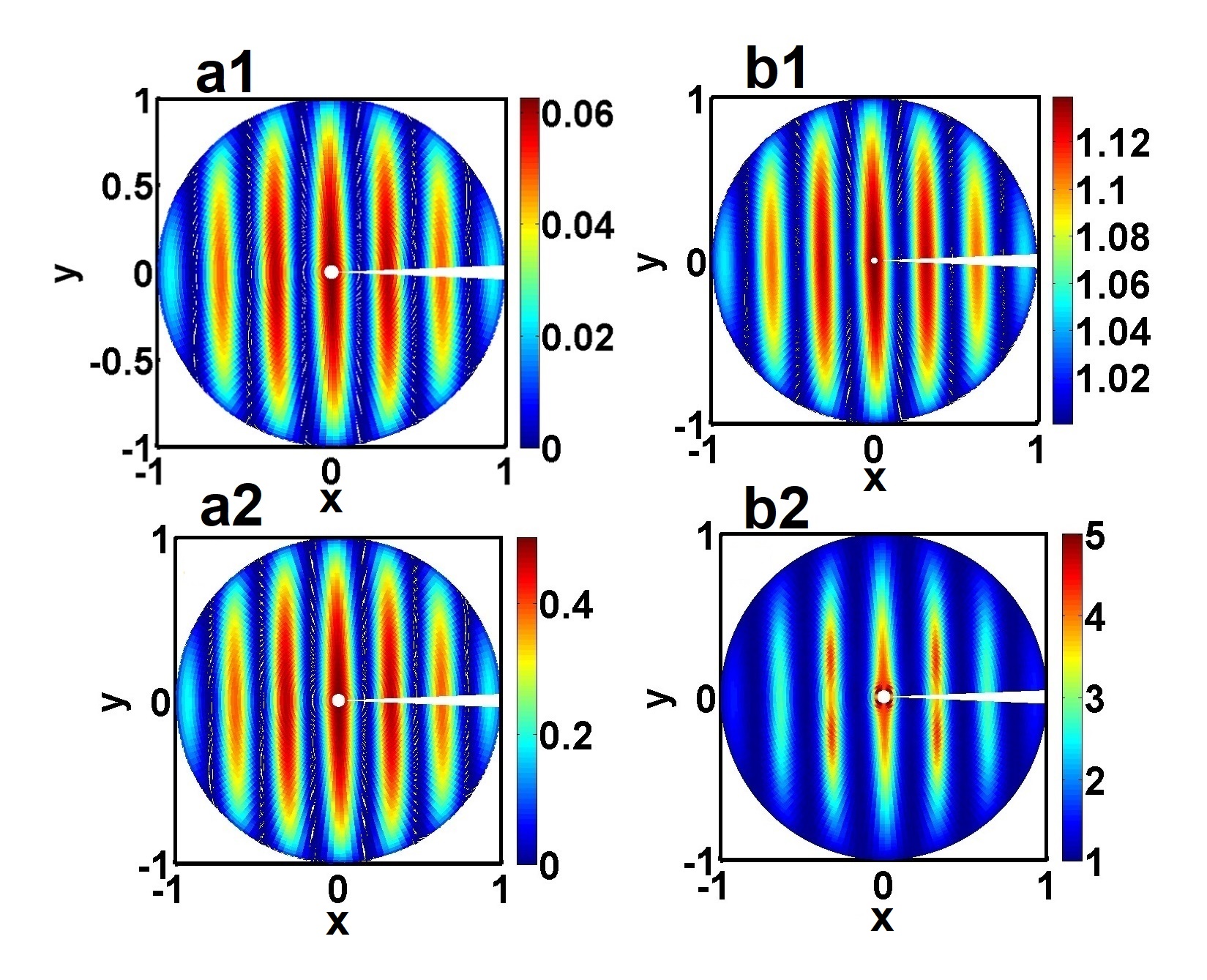}
	\caption{The magnification  calculated for  two solitonic interference lens in x-y plane. The $x$ and $y$ axis are in arbitrary units. We considered the in phase collision($\phi=0$). The observational angle  chosen to be  zero.  (a1)  convergence normalized to $\mathcal{K}=0.01$ (b1) magnification for $\mathcal{K}=0.01$.(a2)  convergence normalized to $\mathcal{K}=0.08$ (b2) magnification for $\mathcal{K}=0.08$. The wavenumber is normalized to the characteristic  wavenumber which is  defined as $k^{\star}=1/r_c$ and we set $k=20$ with the assuming core size $r_c=1$.  }
	\label{magnif}
\end{figure}

\begin{align}
  {\mathcal{K}}=1.1\times 10^{-3}(1-\epsilon) (\frac{10^{-22}\text{eV}}{m})^{2}(\frac{\text{kpc}}{r_{c}})^{3}(\frac{D_{l}}{150Mpc}),
\end{align}
where $\epsilon=D_{l}/ D_{s}$. Assuming that the source is at the distance of $150$Mpc with $\epsilon=1/2$,  the deflection angle from the  potential is as follows
\begin{align}
\vec{\hat{\alpha}}_d=1.45\times 10^{-8} (\frac{10^{-22}\text{eV}}{m})^{2}(\frac{\text{kpc}}{r_{c}})^{2}\nabla{F}.
\end{align}

In Figure \ref{magnif} , we plot the  convergenvce (panel a) and the magnification (panel b) for two solitonic interference lens in x-y plane. We normalize the wavenumbers to $k^\star$ where $k^\star = 1/r_c$. The normalized wavenumber ($k/k^*$) is set to be 20 with the collision phase of $\phi=0$. Also the observational angle  is  zero. 
 In Figure (\ref{magnif}), the panels (a1,b1)  is shown for $\mathcal{K}=0.01$ and (a2,b2) for  $\mathcal{K}=0.08$.
For small value of the convvergence we have the weak lensing approximation, where in the larger ones the approximation $\mu \sim 1+2 \kappa$ does not work and we are in the strong lensing regime.

 \begin{figure}
\centering
\includegraphics[scale=.22]{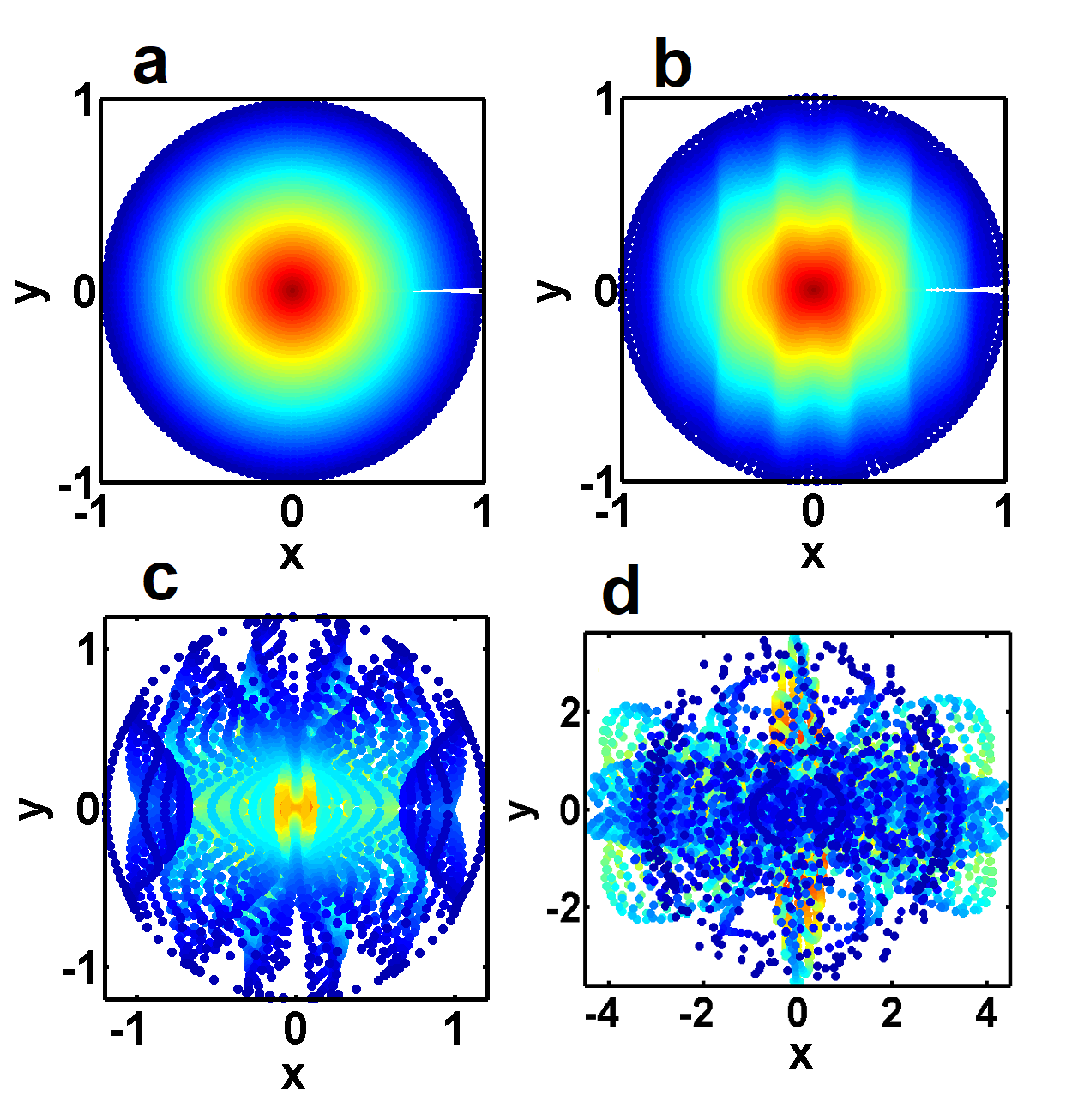}
\caption{ We take a spherically symmetric shape (a panel) as the source in the lensing system. Multiplying this image to the distortion matrix  results in the final images b, c, d with  $\mathcal{K}=0.02$, $\mathcal{K}=0.2$,  $\mathcal{K}=2$ respectively. All parameters are dimensionless. The $x$ and $y$ axis are in arbitrary units. The wavenumber is normalized to the characteristic  wavenumber which is  defined as $k^{\star}=1/r_c$ and we set $k=20$ with the assuming core size $r_c=1$. The  phase of collision is adopted to ($\phi=0$). }
\label{GLImageDistorsion}
  \end{figure}
In Figure \eqref{GLImageDistorsion} we show spherical-symmetric mass density (e.g. elliptical galaxy) which is projected in the  x-y plane and  the distorted image after the effect of gravitational lensing for two colliding FDM. 
 Our aim is to find the footprint of FDM hypothesis with the gravitational lensing methods in dark matter halo cores collision.
In the case of detection of fringes in the $\kappa$, $\gamma_1$ and $\gamma_2$ functions from the weak lensing and measuring the wavelength of the pattern of the fringes, we can measure the Axion's momentum and consequently derive the mass associated to these particles.

 The collision can happen with various phases between the two solitons. In the case of out of phase ($\phi=\pi$) collision,  when the solitonic cores  meet each other, destructive interference occurs by creating a void region  between the two solitons. So for example in a collision of  dark matter halos,  with cores that consist of baryonic matter, this quantum behavior could cause an offset between this two components. as proposed in reference \cite{paredes2016interference}, this may explain the observational  offsets between baryonic matter and dark matter, like  the one that have been observed in  Abell 3827 cluster\cite{taylor2017test,massey2015behaviour}.\\
 We will discuss the observational prospects of this method in section \ref{ObserConsrains}. In upcoming  section we address the evolution of baryonic matter in the colliding solitons gravitational potential as another probe to detect the FDM effects.

  \section{ the  baryonic particles in the colliding fuzzy dark matter core}

  \label{density}
\begin{figure}
\centering
\includegraphics[scale=0.3]{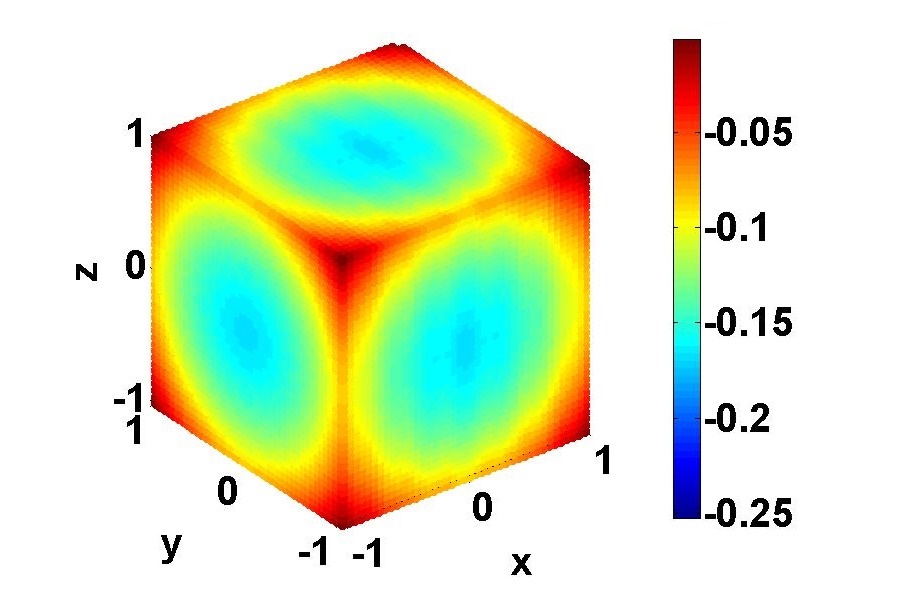}
\caption{The three dimensional distribution of FDM potential  normalized to  ($8\pi G\bar{\rho}(r_c)^2$) and  the  position to $1/k$ which in here we  set $k=20k^*$ and $r_c=1$. 
}
\label{gpotentioa}
\end{figure}

   In this section we investigate, how the baryonic particles would be affected by the fringes that created  in the process of collision of the solitonic part of the fuzzy dark matter. First, we need to find the gravitational potential of the dark matter distribution. Here, we neglect the contribution of the baryonic matters in the gravitational potential as the halo is dominated mainly by the dark matter.
   We use the Poisson equation
     \begin{align}
  \nabla^2 \Phi_t=&4\pi G\rho_{t},
 \label{poisequation}
 \end{align}
where $\rho_{t}$ is given by equation \eqref{rointerf1} for the head on collision. For simplicity we choose the line of collision alone x axis, the analytical solution of equation (\ref{poisequation}) results in
 \begin{align}
   \Phi_t(x,y,z)= -8\pi G\rho_{0} \left(\frac{1}{12}(3r_{c}^2-r^2)+\frac{1}{k^2}\cos(kx+\phi)\right).
 \label{potentioal3D}
 \end{align}
  In Figure \eqref{gpotentioa}, we plot the dimensionless potential $ \Phi_t(x,y,z)/8\pi G\bar{\rho}r_c^2$ 
  for the case of $\phi=0$ phase collision.


For simplicity in calculation,  we separate the potential into a dependent and an independent term with respect to the x parameter as in equation \eqref{potentioal3D}. The one-dimensional potential form equation \eqref{potentioal3D}  is given by
\begin{align}
   \Phi(x)= - 8\pi G\rho_{0}(-\frac{x^2}{12} +\frac{\cos(kx+\phi)}{k^2}).
 \label{intfpotentioal}
 \end{align}
 We note that  the constant term for potential is ignored in redefinition of the  potential.
 In the next part we use this potential to study the dynamics of the baryonic matter in this background. We will investigate  the possibility for the clustering of the baryonic matter under the gravitational potential of the dark matter.
  \subsection{The important time scales in FDM}



 In order to investigate the evolution of baryonic matter during the collision of FDM cores we need to compare the time scales of  the physical events with each other.
We start with the definition of the dynamical time $t_{dyn} = 1/\sqrt{\rho G}$ \cite{padmanabhan1996cosmology} and  substitute the density of core from equation (\ref{cdensity0}),
 \begin{align}
 t_{dyn}=\frac{1}{\sqrt{G\rho}}\simeq 59.2 Myr (\frac{r_{c}}{kpc})^{2}(\frac{m}{10^{-22}ev}).
 \label{t-dyn}
 \end{align}
 Another important time scale is the crossing timescale, $t_{cross}$,
 where the two solitons crossing through each other. By assuming that the collision happens with the relative velocity of $v_{rel}$, the cross time scale  is
 \begin{align}
 t_{cross}=\frac{r_{c}}{v_{rel}}\simeq 9.51 Myr(\frac{r_{c}}{kpc})(\frac{v_{rel}}{100~km/s})^{-1},
 \label{t-cross}
 \end{align}
where the ratio of this two time scale's using equation (\ref{crdensity0}) is
 \begin{align}
 \frac{t_{cross}}{t_{dyn}}\simeq  39(\frac{M_{h}}{10^{12} M_\odot})^{\frac{1}{3}}(\frac{v_{rel}}{100~km/s})^{-1}.
 \label{t-ratio}
 \end{align}
This relation depends on the mass of FDM halo and relative velocity.  The clustering of the baryonic matter happens for the condition of $t_{dyn}<t_{cross}$.
\begin{center}
   \begin{table*}[ht]
  	\centering
  	\label{TimeScalExaample}
  	\caption{The time scales for Two examples of FMD particles mass in which for every mass we supposed two solitonic cores  size (which related to the halo mass). $v_{c}$ is the minimum velocity in which the solitonic  core radius becomes half of particles de broglie wavelength where for the speed lower than this, the cores radius becomes much smaller than the de broglie wavelength of particles and the interference could not appear. We also compute the ${t_{cross}}/{t_{dyn}}$ for each case.}
  	\label{GLconstrains}
  	\vskip 0.1cm
  	\renewcommand*{\arraystretch}{2.1}
  	\begin{tabular}{ |c|c|c|c|c|c|c }
  		\multicolumn{3}{ c } {}\\
  		\cline{1-6}
  	mass of particles $(ev)$&halo mass$(M_\odot)$& $r_c(pc)$ & $v_c(km/s)$&${t_{cross}}/{t_{dyn}}$(for v=400km/s)&${t_{cross}}/{t_{cool}}$(for v=400km/s) \\ \hline
  		
  		\multirow{2}{*}{$m= 10^{-23}$} &$10^{12}$& $1600$ & 400&9.8&1.48 \\ \cline{2-6}
  		&$10^{10}$& $7400$ &84&2&0.32\\ \hline
  		\multirow{2}{*}{$m= 10^{-21}$ } &$10^{12}$&$16$ & 400&9.8&1.48 \\ \cline{2-6}
  		&$10^{10}$& $74$ &84&2&0.32\\ \hline
  	\end{tabular}
  \end{table*}
\end{center}

 The other important parameter is the de broglie wavelength where we can compare this parameter with respect to the size of  cores $r_{c}$ as
 \begin{align}
 \frac{r_{c}}{\lambda_{dB}} \simeq  0.13 (\frac{M_{h}}{10^{12} M_\odot})^{-\frac{1}{3}}(\frac{v_{rel}}{100 km/s}).
 \label{lambdadebro1}
 \end{align}
 This parameter determines where this system has  the quantum or the classical behavior. For the case that $\lambda_{bB}\geq r_{c}$, the systems has quantum behavior, however we can not detect the interference pattern in this condition. Using the of equation  (\ref{lambdadebro1}), we  find a critical velocity where for $v \leq v_{c}$  the quantum behavior is not detectable

 \begin{align}
 v_{c} \sim 400 (\frac{M_{h}}{10^{12} M_\odot})^{\frac{1}{3}}km/s.
 \label{vcrdensity0}
 \end{align}
 This relation shows that for the halo with the size of  Milky Way, collisions with the relative velocities more than  $750km/s$ are in favor of detecting quantum behavior in the collision.
  For the case of   $\lambda_{bB} \leq r_{c}$, we would detect the interference pattern from the quantum behavior, however, for $\lambda_{bB}\ll r_{c}$, due to the limit in the resolution of the instruments, we will detect an envelope for the density pattern and miss detection of the quantum fringes. 



The next time scale that we need to study is the cooling time scales. Assume that the system is consist of a hot gas and the pressure of this gas prevents system from collapsing. Since the pressure is proportional to temperature,  losing  energy with radiation makes the pressure smaller. The cooling process depends on the structure of the gas, metallicity as well as the redshift. We assume a simple case of completely ionized gas of electrons. The cooling time can be written as \cite{binney2011galactic} 
 \begin{eqnarray}
 && t_{cool} = \frac{3}{2}\frac{  k_{B}T(r)}{n(r)\Lambda (T)},
 \label{t-cool}
 \end{eqnarray}
 where $n$ is the number density of particles and  $\Lambda (T)$  is the cooling function
which depends on two physical process of (i) the Bremsstrahlung radiation and (ii)  recombination. Taking the core of galaxies with the dispersion velocity of the order of $\sim 100$~km/s as Milky way galaxy,  the effective temperature would be in the order of kilo-electron volte which is much higher than the recombination energy. So we would expect to have only the Bremsstrahlung radiation from the baryonic plasma with $\Lambda (T) \propto T^{{1}/{2}}$.



For an ideal baryonic gas, assuming power-law function for the profile of the density
 $\rho(r)=\rho_0({r}/{r_0})^{-\alpha}$ and pressure $P(r)=P_0({r}/{r_0})^{-\beta}$ the temperature is obtained as follows
\begin{eqnarray}
T=T_0(\frac{r}{r_0})^{\alpha-\beta} \nonumber\\
\text{and~~~~}
\Lambda (T) \simeq \Lambda_0(\frac{T}{T_0})^{\nu},
\label{temp}
\end{eqnarray}
where $\nu = 1/2$ corresponds to the Bremsstrahlung radiation.
We note that the FDM at the core of halo has a constant density due to the quantum pressure.



For ionized hydrogen in the gravitational potential of FDM, using relation  (\ref{t-cool}) and the virial theorem, we obtain the cooling time as follows
 \begin{eqnarray}
 t_{cool} &\simeq& \frac{2\pi G m_{H}^2 }{3\Lambda (T)}\frac{\rho_{FDM}}{\rho_{H}}r_c^2  \\
&\simeq&1~\text{Myr}(\frac{r_c}{1kpc})^{2}(\frac{10^{-23}erg  s^{-1}cm^{3}}{\Lambda (T)})(\frac{\rho_{FDM}}{100\rho_{H}}). \nonumber
\label{t-bcool}
\end{eqnarray}
The observational evidence from the X-ray emission from the center of galaxies provides that  $\frac{\rho_{FDM}}{100\rho_{H}}<1$ \cite{feldmann2012circum}.

 Now we compute ${t_{cross}}/ t_{cool}$ using the relations (\ref{t-ratio}) and (\ref{t-bcool}),
 \begin{eqnarray}
\frac{t_{cross}}{ t_{cool}} &\simeq&9.51~(\frac{r_c}{1kpc})^{-1}(\frac{10^{-23}erg  s^{-1}cm^{3}}{\Lambda (T)})^{-1}\nonumber\\
&&(\frac{v_{rel}}{100~km/s})^{-1}
(\frac{\rho_{FDM}}{100\rho_{H}})^{-1}.\nonumber\\
\label{tcrossPtcool}
\end{eqnarray}
If the cooling time becomes very small in comparison with the cross-time scale we expect not being able to the detect the clustering of baryonic matter in the quantum fringes of the dark matter. From this relation we can obtain a constrain on the minimum velocity of collision 
\begin{eqnarray}
\frac{v_r}{100km/s} &\geq& 0.60 (\frac{M_{h}}{10^{12} M_\odot})^{\frac{1}{3}}(\frac{10^{-23}erg  sec^{-1}cm^{3}}{\Lambda (T)})^{-1}\nonumber\\
&\times&(\frac{m}{10^{-22}ev}) (\frac{\rho_{FDM}}{100\rho_{H}})^{-1}.
\label{X-rayConstrain}
\end{eqnarray}
This condition also includes the constrain from equation (\ref{vcrdensity0}) .


 Table (I) represents an examples for the case of having FDM particles with two different masses and related host halo dark matter and the core size, respectively. We calculate the ratio of cross time to dynamical time as well. Also, the  values for the minimum relative velocity of the halos satisfying the quantum behavior.
\subsection{The growth of the baryonic  perturbation inside the potential of fuzzy dark matter}
 In this section we study the formation of the baryonic structures within the gravitational potential of FDM. The process of the structure formation can be categorized into two regimes of taking baryonic gas as a dissipative (i.e. $t_{cool}<t_{cross}$) and non-dissipative ( i.e. $t_{cool}>t_{cross}$). In what follows we
 study the formation of the brayonic structures for the non-dissipative condition.

 In the picture of the baryonic structure formation, we study  the rate of  the falling  baryonic matter into the potential given by the equation \eqref{intfpotentioal}. Here, we discuss the growth of the baryonic structure in the linear regime. This framework is suitable for the early stage of the dynamical evolution of baryonic matter.
 We begin with continuity and Euler equations
 \begin{align}
    \frac{\partial \rho_{b}}{\partial t}+\nabla .(\rho_{b} u)=0 ,
        \label{eule}
  \end{align}

   \begin{align}
    \frac{\partial u}{\partial t}+(u.\nabla) . u +\nabla \phi -\frac{\nabla P}{\rho_{b}}=0
    \label{eulerlike}
 \end{align}
 
  \begin{align}
   \nabla^2 \Phi_t=&4\pi G(\rho_{FDM}+\rho_{b}),
  \label {poissionE}
 \end{align}
  the first term of the gravitational potential from equation (\ref{potentioal3D}) is a quadratic term results from the 
 corresponding density of the FDM matter. We call this term as the background term and the static condition for this structure is hold. In this case the gravitational contraction of the FDM and the quantum repulsive forces of the dark halo are equal. Note that we use "zero" subscript for the background terms and "one" subscript for the first order perturbation terms.

The baryonic matter for case that the cooling condition is satisfied can contract in the gravitational potential.  The quadratic term of the background potential (i.e. $\Phi_0 = -3\pi (x^2+y^2+z^2)/4 G\rho_0$) results in a growth of the density of baryonic matter as $\rho_0 \propto t^2$. On the other hand, potential from the ripples (i.e. $\Phi_1$) generates the density contrast of $\delta_{b}=(\rho_{b} - \bar{\rho_{b}})/{\bar{\rho_{b}}}$.
 For the velocity field we adapt zero average velocity field as an initial conditions.
 The pressure of baryonic matter is related to the density as $p=\bar{p}+c_{s}^{2}\delta \rho_{b}$.  Combining the three equations of (\ref{eule}), (\ref{eulerlike}) and (\ref{poissionE}) , results in a differential equation for the evolution of the baryonic density contrast of
 \begin{align}
    \frac{\partial^{2} \delta_{b}}{\partial ^{2}t}-4\pi G \bar{\rho_{b}} \delta_{b}- \nabla^{2} \Phi_{1} -c_{s}^{2}\nabla^{2}\delta_{b} =0.
    \label{perturbEqu}
 \end{align}
We note that according to equation (\ref{potentioal3D}), $\Phi_1$ depends only to the x direction.
We will discuss the solutions in the two conditions of (i) having cooling condition and the pressure term in equation (\ref{perturbEqu}) is smaller than the gravity term (i.e. $c_s\rightarrow 0$) and (ii) there is no cooling and the pressure prevents the condensation of the baryonic matter.

 In the case of ignoring the pressure term compare to the gravity term, the solution for $\delta$ obtain as follows
  \begin{align}
    \delta (x,t)=\left(A e^{\omega t}+B\right) \cos(kx).
    \label{guessans}
 \end{align}
where $  \omega=\sqrt{ 4\pi G \bar{\rho_{b}} - c_{s}^{2}k^{2} }$ and  $B=-{8\pi G\rho_{0}}/{\omega^{2}}$ and  $\rho_{0}$ is the central density of fuzzy dark matter.  We also set the coefficients of $A$ from the boundary conditions. Note that the $k$ is the wave number of fuzzy dark matter which is equal to the wave number of potential of $\Phi_1$.

We adapt $t = 0$ correspond to a time that the density contrast is zero (i.e $\delta (x,t=0)=0$). This condition provides
 $A=-B$. Then the solution of the equation \eqref{perturbEqu} simplifies to
  \begin{align}
    \delta (x,t)= \frac{8\pi G\rho_{0}}{\omega^{2}}(e^{\omega t}-1) \cos(kx),
    \label{dCexpgrows}
 \end{align}
 The solution is the exponential growth of the perturbation term. We note that the solution for the
 perturbation term is valid for $\omega t \ll 1$.

   Now we discuss the regime where the pressure term is larger than the gravity term in equation (\ref{perturbEqu}). Then the solution of this differential equation is
    \begin{align}
    \delta (x,t)= \frac{8\pi G\rho_{0}}{\omega^{2}}\left(1-\cos(\omega t)\right) \cos(kx),
    \label{dCcosgrows}
 \end{align}
  In which ($ \omega=\sqrt{c_{s}^{2}k^{2}-4\pi G \bar{\rho_{b}}} $).
 This is a  periodic density which its amplitude oscillates with the frequency of  $\omega$.  We can call this mode as the acoustic mode for the baryonic matter. The essential condition for the observation of this mode is that the oscillation time scale of the density contrast should be smaller than the collision time scale of the two galaxies, in other word, $t_{osc}<t_{coll}$.

In the next section we will discuss on the observational features of the clumpiness of the baryonic matter within the gravitational potential of the dark matter.


 \subsection{Relaxation mechanism of baryons in collision: Numerical investigation }

 The perturbation approach to describe the evolution of the structures is valid up to the beginning of nonlinear regime (i.e. $\delta <1$). After this phase, we use the numerical method to study the growth of the baryonic structure. In this work, we  use an ensemble of particles of the baryonic matter,  evolving in the potential that created when the two dark matter halo cores are collided and the fringe patterns in the potential is appeared. In this analysis, we assume a non-dissipative system (i.e. $t_{cross} \ll t_{cool}$ or $c_s \neq 0$).



We note that the dark matter dominates in this system and the baryonic particles evolve solely by the dark matter gravitational potential. For simplicity, we use one dimensional potential for the dark matter from 
equation (\ref{intfpotentioal}).
The main concern in this analysis is  studying the  evolution of baryonic particles in this potential.
We examine relaxation time of gas and for the case of $t_{relax}<t_{cross}$, the baryonic structures form within the fringes of the dark matter gravitational potential.





  \begin{figure}
\centering
\includegraphics[scale=0.23]{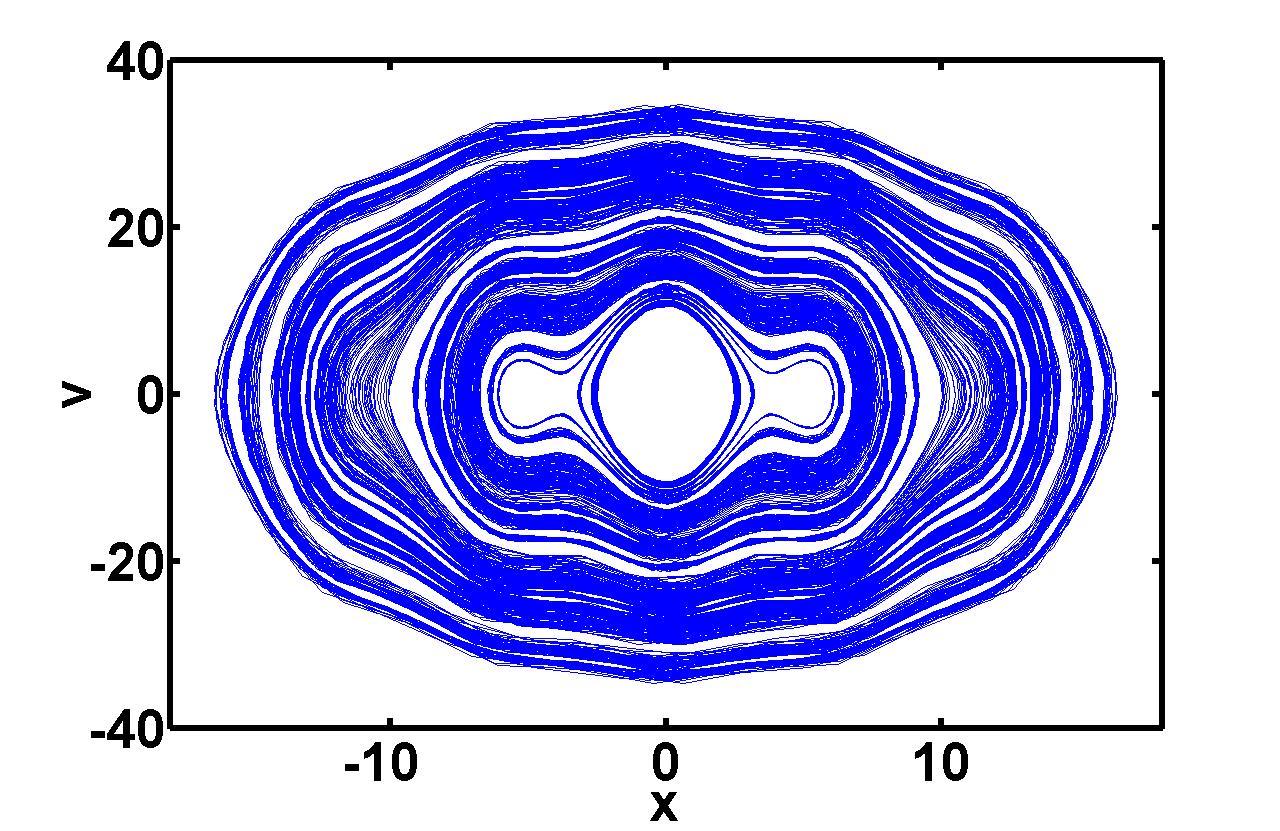}
\caption{ The final stage of particles in the phase space after falling baryonic particles in FDM potential. We use dimensionless coordinates of the velocity and the length by dividing them into $v_{ch}$ and  
$\ell_{ch}$, respectively.}
\label{trajmix}
\end{figure}
For simplicity, we normalize time scale to the dynamical time  $t_{dyn} = 1/\sqrt{G\rho_{0}}$ and length scale with $\ell_{ch} = 1/k$. By diving these two scales, we define a characteristic scale for the velocity as
\begin{align}
v_{ch} = \frac{\ell_{ch}}{t_{dyn}} \simeq 32km/s (\frac{kpc}{r_{c}})^2(\frac{10^{-22}ev}{m})(\frac{100 km/s}{v_{rel}}).
\label{lambdadebro}
\end{align}
In what follows, we normalize time, length and the velocity to the mentioned characteristic scales. 
We also assume the collision is slow, by means that the time scale for crossing is longer than the dynamical time scale (e.g. $t_{cross}>t_{dyn}$). This condition allows the baryonic particles to relax in the gravitational potential of the FDM.

 \begin{figure}
\centering
\includegraphics[scale=0.18]{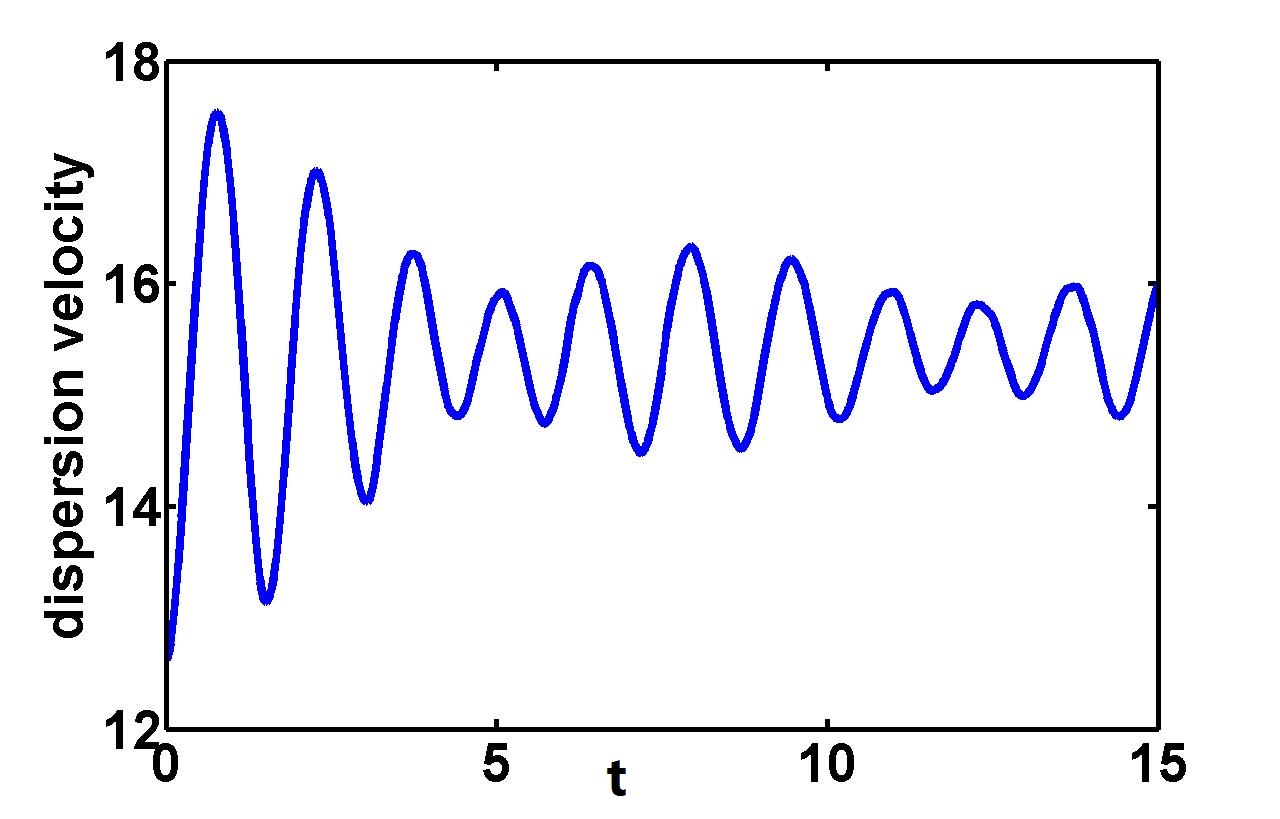}
\caption{Dispersion velocity of the baryonic particles as falling into the gravitational potential of FDM as a function of time where time is normalized to $t_{dyn}$. 
Dispersion velocity approach to a constant value after few dynamical time. 
}
\label{velocityfdmg}
\end{figure}

\begin{figure*}[bt]
	\centering
	\includegraphics[width=1\textwidth]{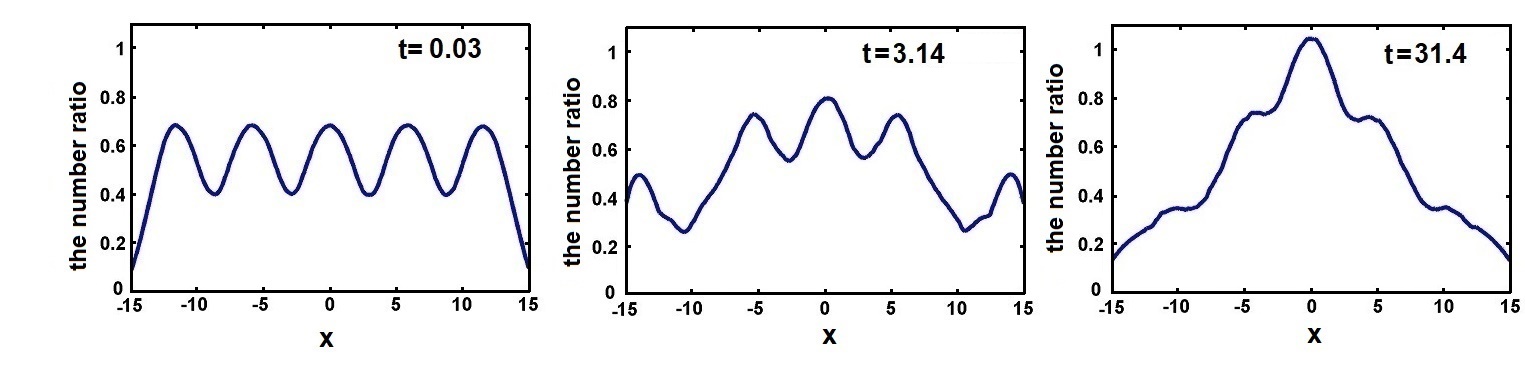}
	\caption{The  contrast in the number of baryonic particles along x axis (we set z and y zero) for the case of $\phi=0$ (in phase collision). We used dimensionless coordinate  by choosing time divided by $1/\sqrt{G\rho_{0}} $, and position by $1/k$ and set $k=1$. The figures  start from initial to the final state. We normalized the number density for all the figures to  maximum value of the final state of FDM model.}
	\label{0profileEvolution}
\end{figure*}

Figure \eqref{trajmix} shows the trajectory of baryonic particles  in the phase space with Maxwelian initial velocity after the particles fall into the gravitational potential generated by FDM. 
 We note that the baryonic particles can heat up due to the collision of the bulges of the two galaxies. As the dynamical time is shorter than the crossing time (see Fig.\ref{velocityfdmg}), the baryonic particles relax in the gravitational ripples by the FDM. From Figure (\ref{velocityfdmg}), the dispersion velocity of the baryonic particles converge to an almost constant value within a few dynamical time-scales. We note that the initial dispersion velocity of baryonic particles due to heating is larger than the virial velocity of the FDM structure (i.e. $v\gg v_{ch}$). The result would be that baryonic particles move through the dark matter potential ripples and spend more times in the ripples.  Figure \eqref{0profileEvolution} demonstrate the schematic distribution of the baryonic matter as a function of time in the gravitational potential of FDM.

Summarizing this part, we have seen that the formation of the baryonic fringes happens in the following condition of $t_{dyn}<t_{relax}<t_{cross}$. Table (I) represents these time-scales for two sets of FDM with different masses.
This condition may satisfy for the cases that two halo are moving through gravitational attraction with a small relative velocity.

\subsection{X-ray observation of colliding FDM}
\label{obs}

Let us assume a volume of gas of baryonic matter with the virial temperature of 
 \begin{align}
 k_{B} T_{vir}\simeq \frac{1}{2}m_{p}v^{2} \Rightarrow T_{vir}\simeq  \frac{m_{p}v^{2}}{2k_{B}},
    \label{VelocityTem}
 \end{align}
where $k_{B}$ is the Boltzmann constant and $m_{p}$ is the baryonic mass of particles.
The main radiation mechanism that causes cooling of the baryonic matter are Bremsstrahlung and recombination of 
ions and free electrons. The radiation rate depends on the square of the density. We can simplify the cooling rate as $n^{2}\Lambda (T)$ where $\Lambda (T)$ is the cooling function. 


The contrast in the X-ray radiation due to cooling is proportional to the density contrast of the baryonic matter as,
\begin{align}
\frac{\delta F}{F} = 2\frac{\delta \rho}{\rho}.
\label{fluxresulution}
\end{align}
It is worth to mention that due to equation \ref{fluxresulution} and Fig.(11) right panel, we know that the flux difference is in the order of the flux. So the flux limit of a detector in the same time can constrain the flux change of FDM effect as well.

For investigating the X-ray images, we refer to our previous results on the 1-Dimension analysis. We can conclude that the three dimensional baryons mass density at the final state  follows the shape of the  FDM  three dimensional potential at the core, so the X-ray radiation can be written as $(n_0 f_\phi(r))^2\Lambda (T)$, in which the $n_0$ is average density of X-ray gas inside galaxies and $f_\phi(r)$ is a dimensionless function which shows the shape of number density profile of X-ray gas and we can calculate with the gravitational potential profile of FDM in equation \eqref{potentioal3D} as, 
\begin{align}
f_\phi(x,y,z)= \frac{45}{(4\pi)^2 r_c^5 }\left(\frac{1}{12}(3r_{c}^2-r^2)+\frac{1}{k^2}\cos(kx+\phi)\right),
\label{xray3D}
\end{align}
where the constant terms is the normalization factor.
We know that the size of X-ray halo is much larger than the size of solitonic core of a galaxy (i.e  $r_{c}\ll r_{g}$ ) and note that the total X-ray intensity received by the observer results from the core and the background radiation from the halo around the galaxy. 

\begin{figure}[h]
	\centering
	\includegraphics[scale=0.4]{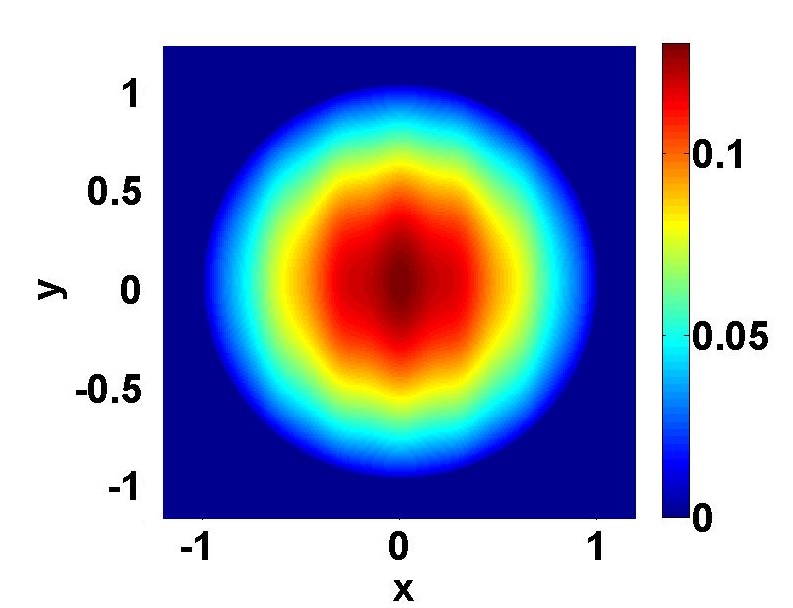}
	\caption{X-ray radiation profile projected along the line of sight.We normalize the length scale to the  core size and assumed wavenumber to be $k=20k^*$ (Note that in this figure $r_c=1$).}
	\label{XrayRadiationProjectedP}
\end{figure}

\begin{figure}[h]
	\centering
	\includegraphics[scale=0.4]{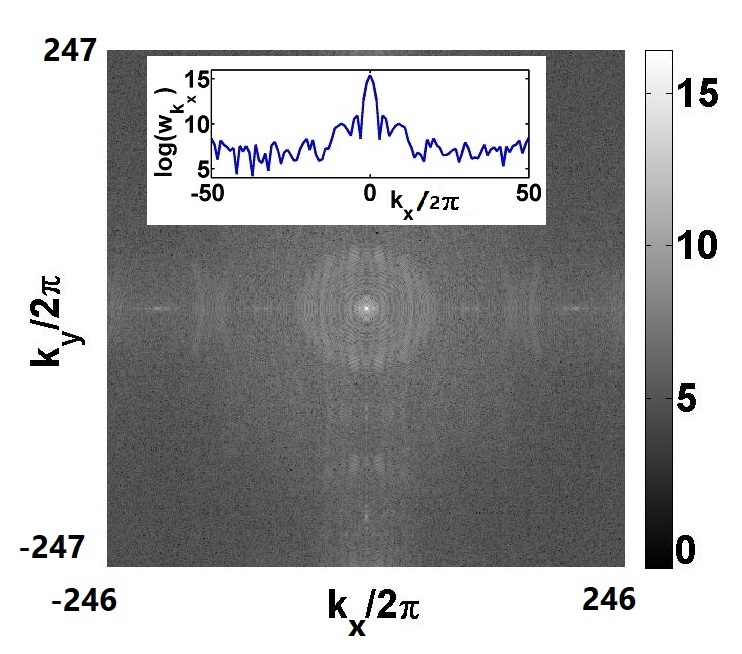}
	\caption{Fourier transform of the projected intensity of the X-ray radiation from Figure (\ref{XrayRadiationProjectedP}). The small panel also represents the spectrum along $x$ axis.}
	\label{FTXrayRadiationProjectedP}
\end{figure}

Integrating the X-ray radiation long the line of sight results in the intensity of X-ray on (x-y) plane as follows
\begin{eqnarray}
{I}_p&=& \int_{-z}^z (n_0 f_\phi(r))^2 \Lambda (T)dz   \\
  &=&(\frac{45 n_0}{(4\pi)^2 r_c^5 })^2[ ( (\frac{1}{6}(3r_{c}^2-x^2-y^2)+\frac{2}{k^2}\cos(kx+\phi))^2    \nonumber \\ 
 & \times & \sqrt{r_{c}^2-x^2-y^2}+\frac{1}{360}(r_{c}^2-x^2-y^2)^{5/2}+ \frac{1}{108}\nonumber \\
 &\times& (r_{c}^2-x^2-y^2)^{3/2}(\frac{1}{12}(3r_{c}^2-x^2-y^2)+\frac{1}{k^2}\cos(kx+\phi))]. \nonumber 
\label{xrayPROJ}
\end{eqnarray}
Figure ($\ref{XrayRadiationProjectedP}$) represents the  X-ray radiation intensity and ripples expected to be detected 
from the colliding FDM cores. In order to study the modes of the ripples, we plot the two dimensional Fourier transform of the image in the Figure ($\ref{FTXrayRadiationProjectedP}$). 
The central part in the Fourier space shown two small dots along $k_x$-axis which is dominant mode from the De Broglie wavelength of the solitons. 
 

\section{Observation prospects}
\label{ObserConsrains}
In this section we discuss on the observability of X-ray ripples from colliding galaxies. 
One of the observational constrain is that the angular resolution of our instrument should be smaller than the angular size of the ripples. Let us assume 
a colliding galaxy at the distance of $D_z$ where the two galaxies with equal mass  have a relative velocity of $v$. The  
angular resolution of the instrument (i.e. $\theta_R$) has to satisfy the following condition between the mass and distance to detect the ripples, 
 \begin{align}
D_z \leq \frac{h}{mv} \frac{1}{\theta_R }.
\label{resolution1}
\end{align}
As  we discussed before according to the equation (\ref{vcrdensity0}) we have a constrain on the relative  velocity of galaxies (i.e. $v>v_c$)  to observe the interference pattern in the collision. Combining equation (\ref{vcrdensity0})  with equation (\ref{resolution1}), we obtain a constrain on the distance of colliding galaxies to detect the interference 
pattern,
 \begin{align}
  \frac{M_{h}}{10^{12} M_\odot} \leq 0.39 (\frac{\theta_{R}}{1 as})^{-3}( \frac{10^{-22}ev}{m})^{3}(\frac{D_z}{100Mpc})^{-3}
 \label{resolution2}
 \end{align}
where $\theta_{R}$ is in arc-seconds.
The Chandra X ray observatory has a resolution of $0.5$ arcsec \cite{chandran}. Accordingly we can search for colliding galaxies with the halo mass of $10^{12}M_\odot$, up to the distance of $\sim 80$ Mpc.





The other observational constrain comes  from the minimum brightness that enable observability of a source. 
Now we assume a soliton core with the radius of  $r_c$ where baryonic matter has th density of $n$ in this state.
 The total radiation intensity  that we receive from this structure is 
\begin{eqnarray}
I_G&\simeq& \frac{\frac{4\pi}{3}r_c^3 {n}^2 \Lambda(T)}{4\pi D_z^2}  .\nonumber\\
&=&4.2 \times 10^{-15}(\frac{n}{1cm^{-3}})^2  (\frac{10^{-23}erg  sec^{-1}cm^{3}}{\Lambda (T)})^{-1}\nonumber\\
&\times&(\frac{M_{h}}{10^{12}M_\odot})^{-1} (\frac{m}{10^{-22}ev})^{-3}(\frac{D_z}{100Mpc})^{-2}   erg sec^{-1}cm^{-2}. \nonumber\\
\end{eqnarray}

The X-ray source should be larger than the instruments sensitivity threshold (i.e. $I_{th}<I_G$) for the observation. This condition puts the constrain on the mass of halo as
\begin{eqnarray}
\label{cons2}
&&\frac{M_{h}}{10^{12} M_\odot} < (\frac{I_{th}}{4\times 10^{-15}erg sec^{-1} cm^{-2} })^{-1} (\frac{n}{cm^{-3}})^{2} \nonumber\\
&\times& (\frac{10^{-23}erg  sec^{-1}cm^{3}}{\Lambda (T)})^{-1}(\frac{m}{10^{-22}ev})^{-3}(\frac{D_z}{100Mpc})^{-2},\nonumber\\
\end{eqnarray}
here we normalized $I_{th}$ to that of the Chandra's threshold. This relation puts an upper limit on the maximum mass of the halo for the detection of quantum ripples.

\begin{figure}
	\centering
	\includegraphics[scale=0.3]{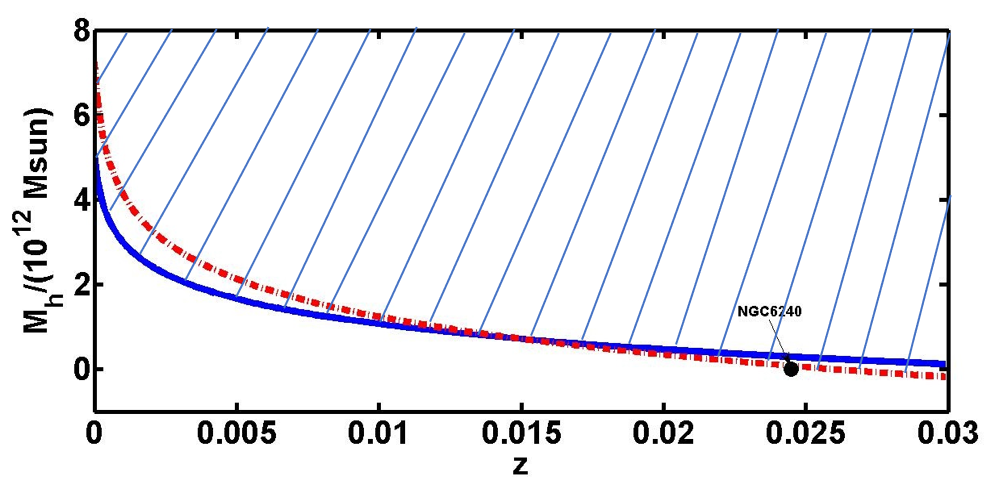}
	\caption{The constrains on the mass of halo as a function of redshift, taking into account the sensitivity of Chandra telescope. Here we adapt 
	$m=10^{-22}ev$, $\Lambda (T)=10^{-23}erg~sec^{-1}cm^{3}$ and  $n=1~cm^{-3}$. The red line represents 
	constrain from the angular resolution of the instrument and the blue line represents limit from the background X-ray. In the hasted area the X-ray fringes are not observable for the Chandra telescope.
	 }
	\label{chanNGC6240}
\end{figure}

Figure (\ref{chanNGC6240})  combines constrains from equation  ($\ref{resolution2}$) and (\ref{cons2}),  representing an upper bound on the observable mass of halo as a function of redshift.  The hashed area excludes the masses 
that can not be observed.


   

\begin{figure}
	\centering
	\includegraphics[scale=0.5]{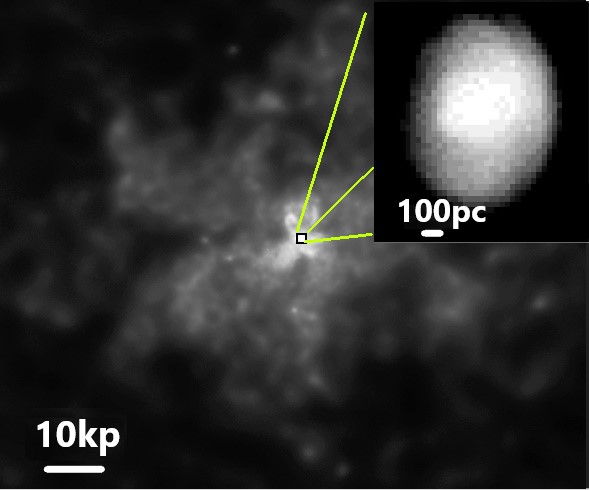}
	\caption{The X-ray image of the colliding galaxies NGC6240\cite{nardini2013exceptional}. }
	\label{NGC6240core}
\end{figure}

\begin{figure}
	\centering
	\includegraphics[scale=0.25]{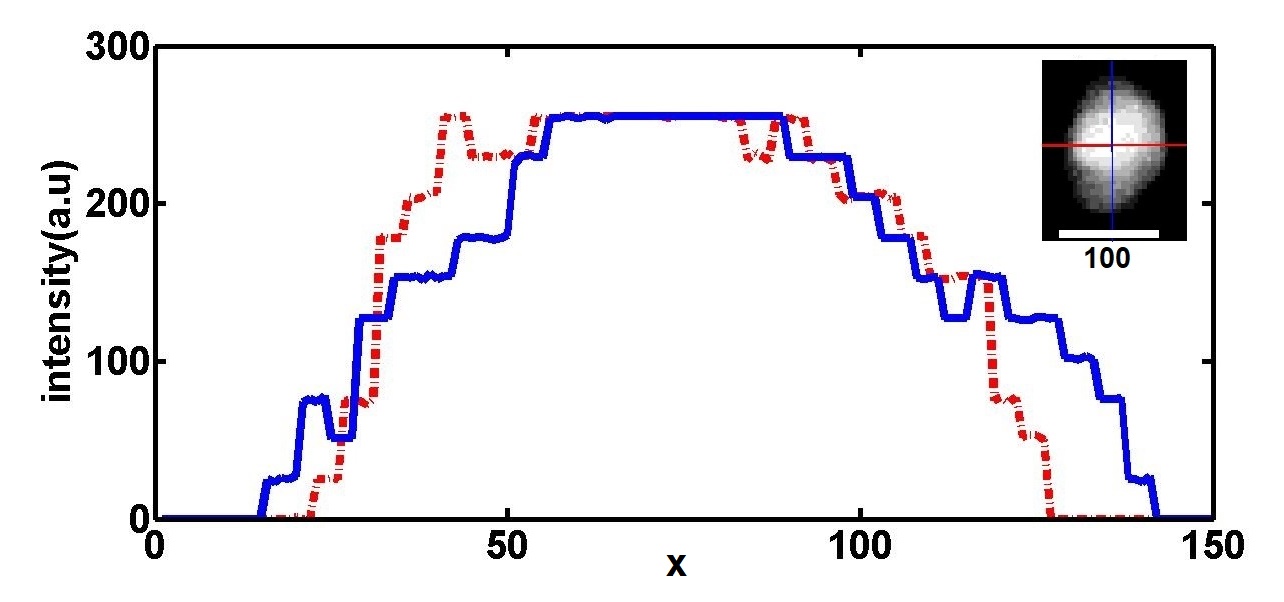}
	\caption{The X-ray intensity of the colliding galaxies NGC6240. Blue solid line shows the X-ray intensity in the vertical line of the inset figure, and the red dashed-dotted line shows the horizontal line \cite{nardini2013exceptional}. }
	\label{XrayINtt6240e}
\end{figure}

\subsection{Searching quantum mechanical  characteristic  in the X-ray images of
	colliding galaxies}

We investigate the X-ray images from the Chandra telescope to identify the colliding galaxies where the cores of the 
galaxies being in the merging stage. We examine NGC 6240 collided system  (\cite{nardini2013exceptional})
for possible detection of X-ray fringes. Figure (\ref{NGC6240core}) shows this colliding system.  Figure (\ref{XrayINtt6240e}) also shows the intensity of the core in $X$ and $Y$ directions around the core of colliding galaxies. For this system the large size of pixels compare to the size of core produces discrete distribution for the intensity of X-ray and with the large size of pixels we can not confirm or rule out the fringes in the X-ray intensity profile. This system has a mass in the order of Milky Way galaxy and at the redshift of $z = 0.024$. Comparing the parameters of system with Figure (\ref{chanNGC6240}), this galaxy is below the resolution of Chandra and we can not resolve the possible fringes in the X-ray profile. In order to investigate the small features in the intensity profile of this system, we need  telescope with higher angular resolution to be smaller than the following value 
 \begin{align}
 \theta_{R} \leq \frac{3.5 \times 10^{-23}ev}{m} as
 \label{ NGC 6240ResolutionConstrain}
 \end{align}
where $m$ is the mass of FDM particles. Accordingly we can assert that  Chandra put a lower limit on FDM mass $\sim 7\times 10^{-23}$eV due to the Eq.( \ref{ NGC 6240ResolutionConstrain}).

 \section{summary and discussion}
\label{conc}
The cold dark matter challenges in  small scale and non-detection of DM particles is a strong motivation to look for beyond the standard model of CDM. In this direction, fuzzy dark matter is one of the favorite models which address more than one of the problems  in the small scales due to its quantum mechanical properties.
One  of the signatures of the quantum mechanical effect of FDM is its interference pattern in the colliding systems.
In this work, we focus on the observation prospect of detecting this type of patterns. We propose that the gravitational lensing effect and the X-ray emission from the distribution of baryonic matter in potential wells of FDM are plausible observational candidates.
 
 
For the case of gravitational lensing we derived all the lensing  parameters for this system and 
discussed how these  parameters depend on the properties of the FDM. In this direction, we proposed that the lensing  convergence maps and the magnification pattern can be used as  observational quantities to detect  the quantum properties of FDM.

We discussed on the clustering of the baryonic matter inside the ripples of the 
gravitational potential  from the collision of the halo cores. 
The fuzzy dark matter sources the gravitational  potential of the Poisson equation, where we studied the evolution of the baryonic matter in the potential configuration of  two solitonic colliding cores.  
The prediction of the linear theory  for the colliding cores is an exponential growth in density contrast and also some ripples on the density field are caused by the baryonic pressure.
Also, we studied the nonlinear evolution  of non-dissipative baryonic matter by tracking the evolution of the test mass  in the mean field of the dark matter potential. The evolution of the ensemble of the particles  results  the clumpiness of the baryonic matter with the same shape of the gravitational potential of FDM. By studying  the evolution of the particles in the phase space, we observed that in the order of $\sim 30$  times of the dynamical time-scale the baryonic fluid relax at the gravitational potential of the dark matter. Finally, we associate a temperature to the  baryonic matter and resulting X-ray radiation for plasma. We  showed that in the X-ray pattern,  we would expect  to detect  a contrast in the intensity which is a straightforward effect of the quantum nature of FDM. 


Finally, we discussed the feasibility of the observations of lensing and X-ray fringes. Accordingly, the threshold of observations due to  resolution limit is discussed for both cases. For the X-ray, we studied the limiting flux as well. 
Combining these two constrains, we obtained an upper limit on the maximum mass of the halos with solitonic cores as function of  redshift, in which the fringes  can be observed in them. 
We have examined the catalog of the colliding galaxies for possible detection of fringes in the X-ray. 
The NGC6240 colliding galaxy at the redshift of $z=0.024$  is a suitable candidate. Our study showed 
no signal from the fringes in the Chandra data and taking into account the angular resolution of 
the Chandra telescope, we  put constrain of $m> 7 \times10^{-23}$ eV on the mass of FDM. 

As the observational prospect, we suggest to observe the nearby colliding galaxies in the X-ray with high resolution instruments.  However, we should note that the simplified model  presented in this work for a colliding galaxies is a first approximation to the problem. More sophisticated models are needed to study the  distribution of the baryons and colliding solitons in this proposal.
Higher resolution observation of colliding system can tighten the constraints on the mass of FDM.
Finally, dedicated observation of lensing is needed by focusing on the colliding galaxies to consider them as a lens. \\ \\



{\bf{Aknowledgements}}\\
 We would like to thank Ali Akbar Abolhasani and Saman Moghimi for useful comments and fruitful discussions.  SB is partially supported by Abdus Salam International Center of Theoretical Physics (ICTP) under
the junior associateship scheme during this work. This research is supported by Sharif
University of Technology Office of Vice President for Research under Grant No. G960202


\bibliographystyle{apsrev}
\bibliography{library}

\end{document}